\documentclass[10pt]{wlscirep}
\usepackage{graphicx, color}

\title{Mutually Exclusive Uncertainty Relations}

\author[1, 2]{Yunlong Xiao}
\author[3, 1 *]{Naihuan Jing}
\affil[1]{School of Mathematics, South China University of Technology, Guangzhou 510640, China}
\affil[2]{Max Planck Institute for Mathematics in the Sciences, Leipzig 04103, Germany}
\affil[3]{Department of Mathematics, North Carolina State University, Raleigh, NC 27695, USA}

\affil[*]{Corresponding author: jing@ncsu.edu}

\begin{abstract}
The uncertainty principle is one of the characteristic properties of quantum theory based on
incompatibility. Apart from the incompatible relation of quantum states, mutually exclusiveness is
another remarkable phenomenon in the information-theoretic foundation of quantum
theory. We investigate the role of mutual exclusive physical states in the recent work of
stronger uncertainty relations for all incompatible observables by Mccone and Pati and
generalize the weighted uncertainty relation
to the product form as well as their multi-observable analogues.
The new bounds capture both incompatibility and mutually exclusiveness, and are tighter compared with the existing bounds.
\end{abstract}
\begin{document}

\flushbottom
\maketitle
%
%
\thispagestyle{empty}

\section*{Introduction}

Heisenberg's uncertainty principle \cite{Heisenberg} is one of the fundamental notions in quantum theory.
The original form is a result of noncommutativity of the position and momentum operators.
Robertson's formulation of Heisenberg's uncertainty principle in matrix mechanics \cite{Robertson} states
that for any pair of observables $A$ and $B$ with bounded spectrums, the product of standard deviations of
$A$ and $B$ is no less than half of the modulus of the expectation value of their commutator:
\begin{equation}\label{e:Robertson}
\Delta A\cdot\Delta B\geqslant\frac{1}{2}|\langle [A, B]\rangle|,
\end{equation}
where $\Delta(A)=\sqrt{\langle A^2\rangle-\langle A\rangle^2}$ is the standard deviation of the self-adjoint
operator $A$. Here the expectation value $\langle \mathcal{O}\rangle=\langle\psi|\mathcal{O}|\psi\rangle$
is over the state $|\psi\rangle$ for any observable $\mathcal{O}$. In fact, Robertson's uncertainty relation
can be derived from a slightly strengthened Schr\"{o}dinger uncertainty inequality \cite{Schrodinger}
\begin{equation}\label{e:schroedinger}
\Delta A^{2}\cdot\Delta B^{2}\geqslant|\frac{1}{2}\langle[A, B]\rangle|^{2}+
|\frac{1}{2}\langle\{\widehat{A}, \widehat{B}\}\rangle|^{2}
\end{equation}
where
$\widehat{\mathcal{O}}=\mathcal{O}-\langle\mathcal{O}\rangle I$.

Besides their importance in quantum mechanics, uncertainty relations play a significant role in
quantum information theory as well \cite{Busch, Lahti, Bialynicki, Deutsch, Wehner, Ghirardi, Bosyk} .
The variance-based uncertainty relations possess clear physical meanings and have a variety of applications in
quantum information processings such as quantum spin squeezing \cite{Walls, Wodkiewicz, Wineland, Kitagawa, Ma} ,
quantum metrology \cite{Giovannetti, Lloyd, Maccone} , and quantum nonlocality \cite{Oppenheim, Li}.

While the early forms of variance-based uncertainty relations are vital to the foundation of quantum theory,
there are two problems still need to be addressed: (\romannumeral1) Homogeneous product of variances may not fully capture the
concept of incompatibility. In other words, a weighted relation may produce a better approximation (e.g.,
the uncertainty relation with R\'{e}nyi entropy \cite{Maassen} and variance-based uncertainty relation for a weighted sum), for more
details and examples, see Ref. \cite{XJLF} ; (\romannumeral2) The existing variance-based uncertainty relations are far from being tight,
and improvement is needed.
One also needs to
know how to generalize the product form to the case of multiple observables for practical applications.

In Ref. \cite{XJLF} , the authors and collaborators have proposed weighted uncertainty relations to answer the first question
and succeeded in improving the uncertainty relation.
Let's recall the weighted uncertainty relation for the sum of variances. For arbitrary two incompatible observables $A$,
$B$ and any real number $\lambda$, the following inequality holds
\begin{equation}
(1+\lambda)\Delta A^{2}+(1+\lambda^{-1})\Delta B^{2}\geqslant\max(\mathcal{L}_{1}, \mathcal{L}_{2})
\end{equation}
with
\begin{align}
\mathcal{L}_{1}(\lambda):=-2i\langle[A, B]\rangle+|\langle\psi|(A-iB)|\psi^{\perp}_{1}\rangle|^{2}
+\lambda^{-1}|\langle\psi|(\lambda A-iB)|\psi^{\perp}_{2}\rangle|^{2}
\end{align}
and
\begin{align}\label{e:L2}
\mathcal{L}_{2}(\lambda):=|\langle\psi|A+B|\psi^{\perp}_{A+B}\rangle|^{2}
+{\lambda}^{-1}|\langle\psi|
(\lambda A-B)|\psi^{\perp}\rangle|^{2},
\end{align}
where $|\psi^{\perp}_{1}\rangle$, $|\psi^{\perp}_{2}\rangle$, $|\psi^{\perp}_{A+B}\rangle$ and $|\psi^{\perp}\rangle$
are orthogonal to $|\psi\rangle$.
In information-theoretic context, it is also natural to quantify the uncertainty by weighted products of variances,
which also help to estimate individual variance as in Ref. \cite{XJLF} .

Recently, Maccone and Pati obtained an amended Heisenberg-Robertson inequality \cite{Pati} :
\begin{equation}\label{e:amended}
\Delta A\Delta B\geqslant\pm\frac{i}{2}\langle[A, B]\rangle/(1-\frac{1}{2}\mid\langle\psi|\frac{A}{\Delta A}\pm i
\frac{B}{\Delta B}|\psi^{\perp}\rangle\mid^{2}),
\end{equation}
which is reduced to Heisenberg-Robertson's uncertainty relation when minimizing the lower bound over $|\psi^{\perp}\rangle$,
and the equality holds at the maximum. This amended inequality gives rise to a stronger uncertainty relation
for almost all incompatible observables, and the improvement is due to the special vector $|\psi^{\perp}\rangle$ perpendicular to the quantum
state $|\psi\rangle$.
We notice that this can be further improved by using the mutually exclusive relation between $|\psi^{\perp}\rangle$ and $|\psi\rangle$.
Moreover, this idea can be generalized to the case of multi-observables. For this reason
the strengthen uncertainty relation thus obtained will be called a {\it mutually
exclusive uncertainty relation}.

The goal of this paper is to answer the aforementioned questions to derive the product form of the
weighted uncertainty relation, and investigate the physical meaning and applications
of the mutual exclusive physical states in variance-based uncertainty relations. Moreover, we will generalize
the product form to multi-observables to give tighter lower bounds.

\section*{Results}

We first generalize the weighted uncertainty relations from the sum form \cite{XJLF} to the product form,
and then introduce {\it mutually exclusive uncertainty relations} (MEUR). After that we derive a couple of lower
bounds based on {\it mutual exclusive physical states} (MEPS), and we show that our results outperform
the bound in Ref. \cite{Pati} , which has been experimentally tested recently \cite{Kun} . Finally,
generalization to multi-observables is also given.

We start with the sum form of the uncertainty relation, which takes equal contribution of the variance from each observable.
However, almost all variance-based
uncertainty relations do not work for the general situation of incompatible observables, and they often
exclude important cases. In Ref. \cite{XJLF} , the authors and collaborators solved this degeneracy problem
by considering weighted uncertainty relations to measure the uncertainty in all cases of incompatible observables.
Using the same idea, we will study the product form of weighted uncertainty relations to give new and alternative
uncertainty relations in the general situation.
The corresponding mathematical tool is the famous {\it Young's inequality}. The new weighted
uncertainty is expected to reveal the lopsided influence from observables. They contain the usual homogeneous
relation of $\Delta A^2\Delta B^2$ as a special case.

\vspace{2ex}
\noindent\textbf{Theorem 1.} {\it Let $A$, $B$ be two observables such that $\Delta A\Delta B>0$,  and $p, q$
two real numbers such that $\frac{1}{p}+\frac{1}{q}=1$. Then the following weighted uncertainty relation
for the product of variances holds.
\begin{align}\label{e:pq}
(\Delta A^2)^{1/p}(\Delta B^2)^{1/q}\geqslant\frac{1}{p}\Delta A^2+\frac{1}{q}\Delta B^2,
\end{align}
where $p<1$, and the equality holds if and only if $\Delta A=\Delta B$. If $p>1$, then $\frac{1}{p}\Delta A^2+\frac{1}{q}\Delta B^2$
becomes a upper bound for the weighted product.}

See Methods for a proof of Theorem 1.

The weighted uncertainty relations for the product of variances have a desirable feature: our measurement of
incompatibility is weighted, which fits well with the reality that
observables usually don't always reach equilibrium, i.e., in physical experiments their contributions may not be the same (cf. Ref. \cite{XJLF}). As
an illustration, let us consider the {\it relative error function} between the uncertainty and
weighted bound, which is defined by
$$f(p)=\frac{(\Delta A^2)^{1/p}(\Delta B^2)^{1/q}-\frac{1}{p}\Delta A^2-\frac{1}{q}\Delta B^2}{(\Delta A^2)^{1/p}(\Delta B^2)^{1/q}}.$$
In general $f$ is a function of both $p$ and $|\psi\rangle$. It is hard to find its extremal points as it involves in partial differential equations. Also
the extremal points hardly occur at homogeneous weights, so incompatible observables
usually don't contribute equally to the uncertainty relation, which explains the need for a weighted uncertainty relation in the
product form.

In what follows, we show how to tighten Maccone and Pati's amended Heisenberg-Robertson
uncertainty relation \cite{Pati} by regarding mutually exclusive physical states as another information resource, and then
generalize the variance-based uncertainty relation to the case of multi-observables.

We will refer to \eqref{e:amended} as a {\it mutually exclusive uncertainty relation} since the states
$|\psi\rangle$ and $|\psi^{\perp}\rangle$ represent two mutual exclusive states in quantum mechanics, which
is the main reason for improving the tightness of the bound.
Next we move further to improve the bound by combining mutually
exclusive relations and weighted relations.

Maccone and Pati's uncertainty relation can be viewed as a singular case in a family
of uncertainty relations parameterized by positive variable $\lambda$, which corresponds to our recent work
on weighted sum of uncertainty relations \cite{XJLF} . We proceed similarly as the case of the amended
Heisenberg-Robertson uncertainty relation by considering a modified square-modulus and Holevo inequalities in
Hilbert space \cite{Holevo} in the following result.

\vspace{2ex}
\noindent\textbf{Theorem 2.} {\it Let $A$ and $B$ be two incompatible observables and $|\psi\rangle$ a fixed
quantum state. Then the mutually exclusive uncertainty relation holds:
\begin{align}\label{e:para}
\Delta A\Delta B\geqslant
\frac{\pm{i}\langle[A, B]\rangle\sqrt{\lambda}}{(1+\lambda)-\mid\langle\psi|
\frac{A}{\Delta A}\pm i\frac{\sqrt{\lambda}B}{\Delta B}|\psi^{\perp}\rangle\mid^{2}}
\end{align}
for any unit vector $|\psi^{\perp}\rangle$ perpendicular to $|\psi\rangle$ and arbitrary parameter
$\lambda>0$.}

See Methods for a proof of Theorem 2.

The obtained variance-based uncertainty relation is stronger than Maccone and Pati's amended uncertainty relation.
In fact, when the maximal value $\mathcal{L}
(\lambda_0, |\psi^{\perp}\rangle)$ is reached at a point $\lambda_0\neq 1$, the new bound is stronger
than that of Maccone-Pati's amended uncertainty relation. Let $\mathcal{L}(\lambda_i, |\psi_i^{\perp}\rangle)$ ($i=1, 2$) be two lower
bounds given in the RHS of (\ref{e:para}), define the {\it tropical sum}
\begin{align}\label{e:improved}
\mathcal{L}(|\psi^{\perp}\rangle)
=\max\{\mathcal{L}(\lambda_1, |\psi^{\perp}\rangle), \mathcal{L}(\lambda_2, |\psi^{\perp}\rangle)\}.
\end{align}
This gives a tighter lower bound when the maximal value of $\mathcal{L}(\lambda_i, |\psi_i^{\perp}\rangle)$
is reached at different direction in $H_{\psi}$ (hyperplane orthogonal to $|\psi\rangle$) for $|\psi_i^{\perp}\rangle$.
In other words, the new lower bound is a piecewise
defined function of MEPS $|\psi^{\perp}\rangle\in H_{\psi}$ taking the maximum of the two bounds.
In particular, for $\lambda_0\neq 1$, the tropical sum $\max\{\mathcal{L}(1,
|\psi^{\perp}\rangle), \mathcal{L}(\lambda_0, |\psi^{\perp}\rangle)\}$ offers a better lower bound than $\mathcal{L}(1, |\psi^
{\perp}\rangle)$, the Maccone-Pati's lower bound. Note that $\mathcal{L}(\lambda, |\psi^{\perp}\rangle)$ may have a smaller
minimum value than $\mathcal{L}(1, |\psi^{\perp}\rangle)$ when $\lambda\neq1$, as $\frac{2\sqrt{\lambda}}{1+\lambda}\leqslant 1$,
while the minimum value of $\mathcal{L}(1, |\psi^{\perp}\rangle)$ is just the bound for Heisenberg-Robertson's uncertainty
relation. Because we only consider the maximum, it does not affect our result.

\begin{figure}
\centering
\includegraphics[width=0.45\textwidth]{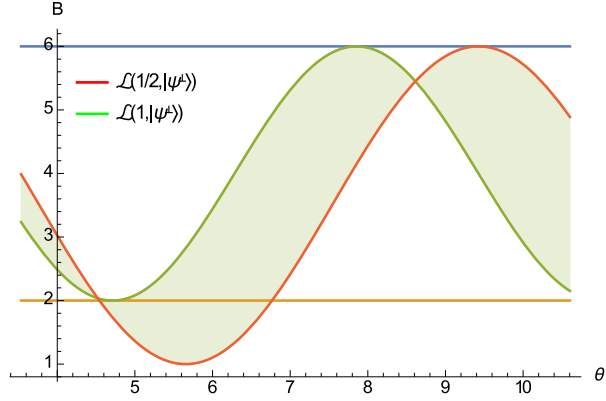}
\caption{Schematic comparison of bounds: Top and middle lines are $\Delta A\Delta B$
and $\pm\frac{i}{2}\langle[A, B]\rangle$ resp. Tropical sum $\max(\mathcal{L}(1, |\psi^{\perp}\rangle),
\mathcal{L}(\frac{1}{2}, |\psi^{\perp}\rangle))$ is the upper boundary above the shadow,
and the red and green
ones are $\mathcal{L}(\frac{1}{2}, |\psi^{\perp}\rangle)$ and Maccone-Pati's bound $\mathcal{L}(1, |\psi^{\perp}\rangle)$ resp.
}
\end{figure}

For example, consider a $4$-dimensional system with state $|\psi\rangle=\cos\frac{\theta}{2}|0\rangle+\sin\frac{\theta}{2}|1\rangle$,
$0\leqslant\theta<\frac{\pi}{2}$ and take the following observables
\begin{align}\label{e:observable}
A=
\left(
\begin{array}{cccc}
  0 & 1 & 0 &0 \\
  1 & 0 & 0 &0 \\
  0 & 0 & 1 &0 \\
  0 & 0 & 0 &-1
\end{array}
\right),
B=
\left(
\begin{array}{cccc}
  1 & -i & 0 &0 \\
  i & -1 & 0 &0 \\
  0 & 0 & 0 &-i \\
  0 & 0 & i &0
\end{array}
\right).
\end{align}
Direct calculation gives
$$\Delta A=\cos\theta, \Delta B=\sqrt{2-\cos^{2}\theta}$$
and
$$\langle A\rangle=\sin\theta,\langle B\rangle=\cos\theta.$$
For $\theta=\frac{\pi}{3}$ and $\lambda=\frac{1}{2}$, set
$$|\psi^{\perp}_{1}\rangle\propto(2-\sqrt{7}+\sqrt{3}i)|0\rangle+(\sqrt{21}-2\sqrt{3}-3i)|1\rangle,$$
and
$$|\psi^{\perp}_{2}\rangle\propto(2-\sqrt{14}+\sqrt{3}i)|0\rangle+(\sqrt{42}-2\sqrt{3}-3i)|1\rangle,$$
both of them have modulus one, then $$\mathcal{L}(1, |\psi^{\perp}_{1})\rangle=\mathcal{L}(\frac{1}{2},
 |\psi^{\perp}_{2})\rangle=\Delta A\Delta B=\frac{\sqrt{7}}{4},$$ meanwhile
$$\mathcal{L}(1, |\psi^{\perp}_{2}\rangle)\approx0.567628<\mathcal{L}(\frac{1}{2}, |\psi^{\perp}_{2})\rangle
=\frac{\sqrt{7}}{4},$$
so
\begin{align}\label{e:max1}
\max(\mathcal{L}(1, |\psi^{\perp}\rangle), \mathcal{L}(\frac{1}{2}, |\psi^{\perp}\rangle))\geq\mathcal{L}(1,
|\psi^{\perp}\rangle).
\end{align}
Both the lower bounds $\max(\mathcal{L}(1, |\psi^{\perp}\rangle), \mathcal{L}(\frac{1}{2}, |\psi^{\perp}\rangle))$
and $\mathcal{L}(1, |\psi^{\perp}\rangle)$ are functions of MEPS $|\psi^{\perp}\rangle$. However, for each
$|\psi^{\perp}\rangle$, $\max(\mathcal{L}(1, |\psi^{\perp}\rangle), \mathcal{L}(\frac{1}{2}, |\psi^{\perp}\rangle))$
gives a better approximation of $\Delta A\Delta B$ than $\mathcal{L}(1, |\psi^{\perp}\rangle)$. FIG. 1 is a schematic
diagram of these two lower bounds. It is clear that $\max_{\lambda}(\mathcal{L}(\lambda, |\psi^{\perp}\rangle)) (\lambda>0)$
provides a closer estimate to $\Delta A\Delta B$:
\begin{align}\label{e:max2}
\Delta A\Delta B\geqslant\max_{\lambda}(\mathcal{L}(\lambda, |\psi^{\perp}\rangle)),
\end{align}
for any unit MEPS $|\psi^{\perp}\rangle$ orthogonal to $|\psi\rangle$. This is due to the fact that the bound
$\max_{\lambda}(\mathcal{L}(\lambda, |\psi^{\perp}\rangle))$ is continuous on both MEPS $|\psi^{\perp}\rangle$ and
$\lambda$, which shows the advantage of our mutually exclusive uncertainty principle. The shadow region in FIG. 2.
illustrates the outline of $\Delta A\Delta B$ and our bound $\max_{\lambda}(\mathcal{L}(\lambda, |\psi^{\perp}\rangle))$.
\begin{figure}
\centering
\includegraphics[width=0.45\textwidth]{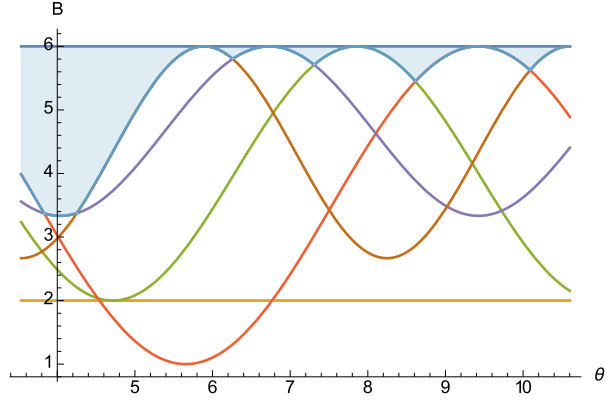}
\caption{Schematic comparison: Top line is $\Delta A\Delta B$. Blue curve is our bound $\max_{\lambda}(\mathcal{L}(\lambda,
|\psi^{\perp}\rangle))$, the shadow region is the difference between $\Delta A\Delta B$ and our bound $\max_{\lambda}
(\mathcal{L}(\lambda, |\psi^{\perp}\rangle))$.
Other bounds are shown in different colors.}
\end{figure}

In FIG. 3, we illustrate our results, showing how the obtained bound
$\max(\mathcal{L}(1, |\psi^{\perp}\rangle), \mathcal{L}(\frac{1}{2}, |\psi^{\perp}\rangle))$ outperforms the recent work of Ref. \cite{Mondal} as well
as the Schr\"{o}dinger uncertainty relation. We consider the angular momenta $L_{x}$ and $L_{y}$ for spin-$1$ particle with state
$|\psi\rangle=\cos\theta|1\rangle-\sin\theta|0\rangle$ and $|\psi^{\perp}\rangle=\sin\theta|1\rangle+\cos\theta|0\rangle$, where $|0\rangle$ and $|1\rangle$ are eigenstates of the angular momentum $L_{z}$.

\begin{figure}
\centering
\includegraphics[width=0.45\textwidth]{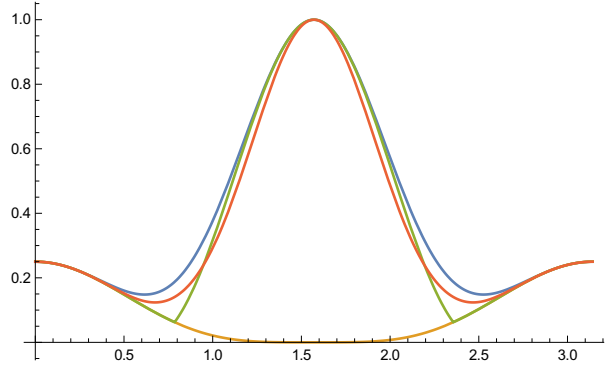}
\caption{Lower bounds of $\Delta A^{2}\Delta B^{2}$ for a family of spin-$1$ particles: $\Delta A^{2}\Delta B^{2}$, our bound
 $\max(\mathcal{L}(1, |\psi^{\perp}\rangle), \mathcal{L}(\frac{1}{2}, |\psi^{\perp}\rangle))$, the bound given
 by Eq. (3) in Ref. \cite{Mondal} and the Schr\"{o}dinger bound are respectively shown in blue, orange, green and yellow.}
\end{figure}

Mutually exclusive physical states with different directions in $H_{\psi}$ offer different kinds of mutually
exclusive information and improvement of the uncertainty relation. When such an experiment of the mutually exclusive uncertainty relation is
performed, one is expected to have infinitely many strong lower bounds
of the variance-based uncertainty relation.

Now we further generalize the uncertainty relations to multi-observables.
For simplicity, write
$$\mathcal{L}(\lambda, |\psi^{\perp}\rangle):=\pm\frac{i}{2}\langle[A, B]\rangle f(\lambda,|\psi^{\perp}\rangle; A, B).$$

So
\begin{align}\label{e:f}
f(\lambda, |\psi^{\perp}\rangle; A, B)
=\frac{2\sqrt{\lambda}}{(1+\lambda)-\mid\langle\psi|\frac{A}{\Delta A}
\pm i\frac{\sqrt{\lambda}B}{\Delta B}|\psi^{\perp}\rangle\mid^{2}}
\end{align}
is continuous on both MEPS $|\psi^{\perp}\rangle$ and $\lambda$. Repeatedly using  (\ref{e:para}) for $|\psi^{\perp}_{jk}
\rangle$ and $\lambda_{jk}$, we obtain the following relation.

\vspace{2ex}
\noindent\textbf{Theorem 3.} {\it Let $A_{1}$, $A_{2}$, $\ldots$, $A_{n}$ be $n$ incompatible observables, $|\psi\rangle$
a fixed quantum state and $\lambda_{jk}$ positive real numbers, we have
\begin{align}\label{e:multi}
\Delta A_{1}\Delta A_{2}\cdots\Delta A_{n}
\geqslant(-\frac{i}{2})^{\frac{n}{2}}\max\limits_{\lambda_{jk}}[\prod\limits_{j>k}(f(\lambda_{jk}, |\psi^{\perp}_{jk}\rangle;
A_{j}, A_{k}))\langle[A_{j}, A_{k}]\rangle]^{\frac{1}{n-1}},
\end{align}
for any MEPS $|\psi^{\perp}_{jk}\rangle$ orthogonal to $|\psi\rangle$ with modulus one.
If some $-\frac{i}{2}\langle[A_{j}, A_{k}]\rangle$ is negative,
a negative sign is inserted into the RHS of   (\ref{e:multi}) to ensure positivity.
The equality holds if and only if MEPS $|\psi^{\perp}_{jk}\rangle\propto(\frac{\widehat{A}_{j}}{\Delta A_{j}}\mp i\frac{\sqrt
{\lambda_{jk}}\widehat{A}_{k}}{\Delta A_{k}})|\psi\rangle$ for all $j>k$.}

As a corollary, Theorem 3 leads to a simple bound of the uncertainty relation for multi-observables.

\vspace{2ex}
\noindent\textbf{Corollary 1.} {\it Let $A_{1}$, $A_{2}$, $\ldots$, $A_{n}$ be $n$ incompatible observables, then the following
uncertainty relation holds}
\begin{align}\label{e:multic}
\Delta A_{1}\Delta A_{2}\cdots\Delta A_{n}
\geqslant(-\frac{i}{2})^{\frac{n}{2}}[\prod\limits_{j>k}\langle [A_{j}, A_{k}]\rangle]^{\frac{1}{n-1}}.
\end{align}

See Methods for a proof of Corollary 1.

Next, we provide yet another mutually exclusive uncertainty relation.

\begin{figure}
\centering
\includegraphics[width=0.45\textwidth]{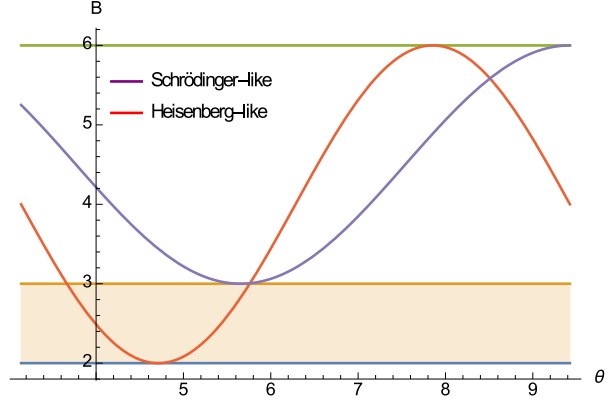}
\caption{Schematic comparison of uncertainty relations: Top, middle and bottom lines are
$\Delta A^{2}\Delta B^{2}$, Schr\"{o}dinger's and the square of Heisenberg's bounds resp.
Orange and purple curves are the square of Maccone-Pati's amended Heisenberg bound and our amended Schr\"{o}dinger
bound resp.}
\end{figure}

\vspace{2ex}
\noindent\textbf{Theorem 4.} {\it Let $A$ and $B$ be two incompatible observables and $|\psi\rangle$ a fixed quantum state. Then
\begin{align}\label{e:hh}
\Delta A^{2}\Delta B^{2}
\geqslant\frac{(g(|\psi^{\perp}_{1}\rangle, |\psi^{\perp}_{2}\rangle)+2s)+\mid g(|\psi^{\perp}_{1}\rangle, |\psi^{\perp}_{2}\rangle)-2s\mid}{4},
\end{align}
for any unit MEPS $|\psi^{\perp}_{i}\rangle (i=1, 2)$ orthogonal to $|\psi\rangle$, with
\begin{align}\label{e:g}
g(|\psi^{\perp}_{1}\rangle, |\psi^{\perp}_{2}\rangle)
=\frac{\mid\frac{1}{2}\langle[A, B]\rangle\mid^{2}}{(1-\frac{1}{2}\mid\langle|\psi|\frac{A}{\Delta A}+i\frac{B}
{\Delta B}|\psi^{\perp}_{1}\rangle\mid^{2})^{2}}
+\frac{\mid\frac{1}{2}\langle\{A, B\}\rangle-\langle A\rangle\langle B\rangle\mid^{2}}{(1-\frac{1}{2}\mid\langle
|\psi|\frac{A}{\Delta A}+\frac{B}{\Delta B}|\psi^{\perp}_{2}\rangle\mid^{2})^{2}},
\end{align}
where MEPS $|\psi_i^{\perp}\rangle$ are unit vectors in $H_{\psi}$.}

See Methods for a proof of Theorem 4.

Obviously, (\ref{e:hh}) can be seen as an {\it amended Schr\"{o}dinger inequality} and also
offers a better bound than (\ref{e:schroedinger}) and Maccone-Pati's relation (\ref{e:amended}). FIG. 4 illustrates the schematic comparison.

In general, if there exists an operator $M$ for $A$ and $B$ such that $\langle M\rangle=0$, $\langle MM^{\dag}\rangle=2\pm2\frac{\langle\widehat{A}\widehat{B}\rangle}{\Delta A\Delta B}$, then we have the following:

\vspace{2ex}
\noindent\textbf{Remark 1.} {\it Let $A$ and $B$ be two incompatible observables and $|\psi\rangle$ a fixed quantum state.
We claim the following mutually exclusive uncertainty relation holds:}
\begin{align}\label{e:amended2}
\Delta A^{2}\Delta B^{2}\geqslant\frac{\mid\frac{1}{2}\langle[A, B]\rangle\mid^{2}+
\mid\frac{1}{2}\langle\{\widehat{A}, \widehat{B}\}\rangle|^2}{(1-\frac{1}{2}\mid\langle\psi|M|\psi^{\perp}\rangle\mid^{2})^{2}}.
\end{align}
Eq. (\ref{e:amended2}) also gives a generalized Schr\"{o}dinger uncertainty relation. Here as usual MEPS $|\psi^{\perp}\rangle$ is
any unit vector perpendicular to $|\psi\rangle$. The proof of Theorem 4 and Remark 1 are similar to that
of Theorem 2, so we sketch it here. It is easy to see that the RHS of   (\ref{e:amended2}) reduces to the
lower bound of Schr\"{o}dinger's uncertainty relation (\ref{e:schroedinger}) when minimizing over $|\psi^{\perp}\rangle$, and the
equality holds at the maximum. The corresponding uncertainty relation for arbitrary $n$ observables is the following result.

\vspace{2ex}
\noindent\textbf{Theorem 5.} {\it Let $A_{1}$, $A_{2}$, $\ldots$, $A_{n}$ be $n$ incompatible observables, $|\psi\rangle$
a fixed quantum state and $\lambda_{jk}$ positive real numbers. Then we have that
\begin{align}\label{e:amended3}
\Delta A^{2}_{1}\Delta A^{2}_{2}\cdots\Delta A^{2}_{n}
\geqslant[\prod\limits_{j>k}\frac{\mid\frac{1}{2}\langle[A_{j}, A_{k}]\rangle\mid^{2}+\mid\frac{1}{2}\langle\{\widehat{A}_{j}, \widehat{A}_{k}\}\rangle\mid^{2}}{(1-\frac{1}{2}\mid\langle\psi|M_{jk}|\psi^{\perp}_{jk}\rangle\mid^{2})^{2}}]^{\frac{1}{n-1}},
\end{align}
where $M_{jk}$ satisfy $\langle M_{jk}\rangle=0$, $\langle M_{jk}M_{jk}^{\dag}\rangle=2\pm2\frac{\langle\widehat{A_{j}}\widehat
{A_{k}}\rangle}{\Delta A_{j}\Delta A_{k}}$ and MEPS $|\psi^{\perp}_{jk}\rangle$ orthogonal to $|\psi\rangle$ with modulus one.}

The RHS of (\ref{e:amended3}) has the minimum value
$$[\prod\limits_{j>k}(\mid\frac{1}{2}\langle[A_{j}, A_{k}]\rangle\mid^{2}+\mid\frac{1}{2}\langle\{\widehat{A}_{j}, \widehat{A}_{k}\}\rangle\mid^{2})]^{\frac{1}{n-1}}$$
and the equality holds at the maximum. Therefore one obtains the following corollary.

\vspace{2ex}
\noindent\textbf{Corollary 2.} {\it Let $A_{1}$, $A_{2}$, $\ldots$, $A_{n}$ be $n$ incompatible observables, then the following
uncertainty relation holds}
\begin{align}\label{e:multic}
\Delta A_{1}\Delta A_{2}\cdots\Delta A_{n}
\geqslant[\prod\limits_{j>k}(\mid\frac{1}{2}\langle[A_{j}, A_{k}]\rangle\mid^{2}+\mid\frac{1}{2}\langle\{\widehat{A}_{j}, \widehat{A}_{k}\}\rangle\mid^{2})]^{\frac{1}{n-1}}.
\end{align}

See Methods for a proof of Corollary 2.

We note that our enhanced Schr\"odinger uncertainty relations offer significantly tighter lower bounds than that of
Maccone-Pati's uncertainty relations for multi-observables, as our lower bound contains an extra term of $|\frac 12
\langle \{\widehat{A}, \widehat{B}\}\rangle|^2$ (compare   (\ref{e:Robertson}) with   (\ref{e:schroedinger})).

Finally, we remark that we can also replace the non-hermitian operator $\frac{A}{\Delta A}\pm i\frac{B}{\Delta B}$ in
  (\ref{e:amended}) by a hermitian one. A natural consideration is the amended uncertainty relation
\begin{align}\label{e:amended4}
\Delta A\Delta B\geqslant\frac{\frac{1}{2}\mid\langle\{A, B\}\rangle-\langle A\rangle\langle B\rangle\mid}{1-\frac{1}{2}\mid
\langle|\psi|\frac{A}{\Delta A}+\frac{B}{\Delta B}|\psi^{\perp}\rangle\mid^{2}},
\end{align}
for any unit MEPS $|\psi^{\perp}\rangle$ perpendicular to $|\psi\rangle$. The corresponding uncertainty relation for
multi-observables can also be generalized.

The minimum of Maccone and Pati's amended bound $\mathcal{L}(1, |\psi^{\perp}\rangle)$ in the RHS of   (\ref{e:amended})
agrees with the bound in Heisenberg-Robertson's uncertainty relation, which is weaker than Schr\"{o}dinger's bound in
  (\ref{e:schroedinger}). We point out that the bound given as a continuous function of MEPS's
will always produce a better lower bound. In fact, the continuity of $\mathcal{L}(1, |\psi^{\perp}\rangle)$ in MEPS
$|\psi^{\perp}\rangle$ implies that there exists suitable $|\psi^{\perp}_{0}\rangle$ such that $\mathcal{L}(1,
|\psi^{\perp}_{0}\rangle)$ is tighter than the bound of Heisenberg-Robertson's uncertainty relation. Similarly our
lower bound given in   (\ref{e:h}) or more generally in   (\ref{e:amended2}) provides a tighter lower bound than
the enhanced Schr\"odinger's uncertainty relation (\ref{e:schroedinger}). This shows the advantage of lower bounds
with MEPS's.
Furthermore, lower bounds with more variables give better estimates for the
product of variances of observables, as in   (\ref{e:amended3}).

\section*{Conclusions}

The Heisenberg-Robertson uncertainty relation is a fundamental principle of quantum theory. It has been recently
generalized by Maccone and Pati to an enhanced uncertainty relation for two observables via mutually exclusive
physical states.
Based on these and weighted uncertainty relations \cite{XJLF} , we have derived uncertainty relations for the product of variances from mutually
exclusive physical states (MEPS) and offered tighter bounds.

In summary, we have proposed generalization of variance-based uncertainty relations. By virtue of MEPS, we have introduced a family of infinitely
many Schr\"odinger-like uncertainty relations with tighter lower bounds for the product of variances. Indeed,
our mutually exclusive uncertainty relations can be degenerated to the classical variance-based uncertainty relations by fixing MEPS and the weight.
Also, our study further shows that the mutually exclusiveness between states is a promising information resource.

\section*{Methods}

\noindent\textbf{Proof of Theorem 1.} To prove the theorem, we recall {\it Young's inequality} \cite{Zorich} : for $\frac{1}{p}+\frac{1}{q}=1$,
$p<1$ one has that
\begin{align}\label{e:pq1}
(\Delta A^2)^{1/p}(\Delta B^2)^{1/q}\geqslant\frac{1}{p}\Delta A^2+\frac{1}{q}\Delta B^2.
\end{align}
Note that the right-hand side (RHS) may be
negative if $p<1$. But this can be avoided by using the symmetry of Young's inequality to
get
$$(\Delta A^2)^{1/q}(\Delta B^2)^{1/p}\geqslant\frac{1}{q}
\Delta A^2+\frac{1}{p}\Delta B^2>0.$$
Thus our bound is nontrivial. We remark that if $p>1$, it is directly from the {\it Young's inequality} \cite{Zorich}
\begin{align}\label{e:pq2}
(\Delta A^2)^{1/p}(\Delta B^2)^{1/q}\leqslant\frac{1}{p}\Delta A^2+\frac{1}{q}\Delta B^2,
\end{align}
and equality holds in (\ref{e:pq1}) and (\ref{e:pq2}) only when $\Delta A=\Delta B$.$\blacksquare$
\vspace{2ex}

\noindent\textbf{Proof of Theorem 2.} Here we provide two proofs of the proposed mutually exclusive uncertainty relation \eqref{e:para}.
The first one, based on
weighted relations \cite{XJLF} , is a natural deformation of Ref. \cite{Pati} and is sketched as follows. By maximizing the RHS of (\ref{e:para}), we see that the maximum $\Delta A\Delta B$ is achieved
when the {\it mutually exclusive physical state} (MEPS) $|\psi^{\perp}\rangle\propto(\frac{\widehat{A}}{\Delta A}\mp i\frac
{\sqrt{\lambda}\widehat{B}}{\Delta B})|\psi\rangle$. Clearly
our uncertainty relation contains (\ref{e:amended}) as a special case of $\lambda=1$.

The second proof uses geometric property and is preferred
because of its mathematical simplicity and also working for the amended Heisenberg-Robertson uncertainty relation
\cite{Pati} .  In fact, the RHS of \eqref{e:amended}, denoted by $\mathcal{L}
(\lambda, |\psi^{\perp}\rangle)$, is a continuous function of $\lambda$ and the unit MEPS
$|\psi^{\perp}\rangle$. By the vector projection, the maximum value
$\Delta A\Delta B$ of $\mathcal{L}(\lambda, |\psi^{\perp}\rangle)$ over the hyperplane of $|\psi^{\perp}\rangle$ is attained when $|\psi^{\perp}
\rangle\propto(\frac{\widehat{A}}{\Delta A}\mp i\frac{\sqrt{\lambda}\widehat{B}}{\Delta B})|\psi\rangle$.
 Therefore for any $\lambda>0$
$$\max_{|\psi^{\perp}\rangle}\mathcal{L}(\lambda, |\psi^{\perp}\rangle)=\max_{|\psi^{\perp}\rangle}
\mathcal{L}(1, |\psi^{\perp}\rangle)=\Delta A\Delta B,$$
where $\mathcal{L}(1, |\psi^{\perp}\rangle)$ is the RHS of   (\ref{e:amended}).
Similarly
\begin{align*}
\min_{|\psi^{\perp}\rangle}\mathcal{L}(\lambda, |\psi^{\perp}\rangle)=
\frac{\pm{i}\langle[A, B]\rangle\sqrt{\lambda}}{1+\lambda}\leqslant\frac{\pm{i}\langle[A, B]\rangle}{2},
\end{align*}
for any $\lambda>0$ and the equality holds if $\lambda=1$, which implies   (\ref{e:para}) and
completes the second proof.$\blacksquare$
\vspace{2ex}

\noindent\textbf{Proof of Corollary 1.} Obviously, taking the minimum of (\ref{e:multi}) over MEPS $|\psi^{\perp}_{jk}\rangle$ implies that
\begin{align}\label{e:multi2}
\Delta A_{1}\Delta A_{2}\cdots\Delta A_{n}
\geqslant&(-\frac{i}{2})^{\frac{n}{2}}\max\limits_{\lambda_{jk}}[\prod\limits_{j>k}
\frac{2\sqrt{\lambda_{jk}}}{1+\lambda_{jk}}\langle [A_{j}, A_{k}]\rangle]^{\frac{1}{n-1}}\\
=&(-\frac{i}{2})^{\frac{n}{2}}[\prod\limits_{j>k}\langle [A_{j}, A_{k}]\rangle]^{\frac{1}{n-1}}. \notag
\end{align}
When $\lambda_{jk}=1$ for all $j>k$, the minimum is $(-\frac{i}{2})^{\frac{n}{2}}[\prod\limits_{j>k}\langle[A_{j}, A_{k}]
\rangle]^{\frac{1}{n-1}}$. Meanwhile if $\lambda_{jk}$ and MEPS $|\psi^{\perp}_{jk}\rangle$ vary, Eq. (\ref{e:multi})
provides a family of mutually exclusive uncertainty relations for arbitrary $n$ observables with   (\ref{e:multi2})
as the lower bound.$\blacksquare$
\vspace{2ex}

\noindent\textbf{Proof of Theorem 4.}
By the same method used in deriving
(\ref{e:para}) it follows that $\max_{|\psi_i^{\perp}\rangle} g=2\Delta A^{2}\Delta B^{2}$, and
$\min_{|\psi_i^{\perp}\rangle} g$ is
\begin{align}\label{e:s}
s=\mid\frac{1}{2}\langle[A, B]\rangle\mid^{2}+\mid\frac{1}{2}\langle\{A, B\}\rangle-\langle A\rangle\langle B\rangle\mid^{2},
\end{align}
which equals to the lower bound of the Schr\"odinger uncertainty (\ref{e:schroedinger}).
We can modify $g$ into a function with the same maximum and lower bound as Schr\"{o}dinger's uncertainty relation.
Note that $s\leqslant \Delta A^{2}\Delta B^{2}$, then
\begin{align}\label{e:scaling}
\Delta A^{2}\Delta B^{2}\geqslant(g(|\psi^{\perp}_{1}\rangle, |\psi^{\perp}_{2}\rangle)-s)\frac{\Delta A^{2}\Delta B^{2}-s}
{2\Delta A^{2}\Delta B^{2}-s}+s,
\end{align}
which is equivalent to (by solving $\Delta A^{2}\Delta B^{2}$)
\begin{align}\label{e:h}
\Delta A^{2}\Delta B^{2}
\geqslant\frac{(g(|\psi^{\perp}_{1}\rangle, |\psi^{\perp}_{2}\rangle)+2s)+\mid g(|\psi^{\perp}_{1}\rangle, |\psi^{\perp}_{2}\rangle)-2s\mid}{4},
\end{align}
for any unit MEPS $|\psi^{\perp}_{i}\rangle (i=1, 2)$ orthogonal to $|\psi\rangle$. In fact, let $h(|\psi^{\perp}_{1}\rangle, |\psi^{\perp}_
{2}\rangle)$ be the RHS of (\ref{e:h}). It is easy to see that $\max_{|\psi^{\perp}_i\rangle}h=\Delta A^{2}\Delta B^{2}$ and
$$\min_{|\psi^{\perp}_i\rangle}h=\mid\frac{1}{2}\langle[A, B]\rangle\mid^{2}+\mid\frac{1}{2}\langle\{\hat{A}, \hat{B}\}\rangle\mid^2=s.$$
Hence we have the mutually exclusive uncertainty relation appeared in (\ref{e:h}).$\blacksquare$
\vspace{2ex}

\noindent\textbf{Proof of Corollary 2.} Apparently, taking the minimum of \eqref{e:amended3} over MEPS $|\psi^{\perp}_{jk}\rangle$ implies that
\begin{align}\label{e:multi4}
\Delta A^{2}_{1}\Delta A^{2}_{2}\cdots\Delta A^{2}_{n}
\geqslant&[\prod\limits_{j>k}\frac{\mid\frac{1}{2}\langle[A_{j}, A_{k}]\rangle\mid^{2}+\mid\frac{1}{2}\langle\{\widehat{A}_{j}, \widehat{A}_{k}\}\rangle\mid^{2}}{(1-\frac{1}{2}\mid\langle\psi|M_{jk}|\psi^{\perp}_{jk}\rangle\mid^{2})^{2}}]^{\frac{1}{n-1}}\\
=&[\prod\limits_{j>k}(\mid\frac{1}{2}\langle[A_{j}, A_{k}]\rangle\mid^{2}+\mid\frac{1}{2}\langle\{\widehat{A}_{j}, \widehat{A}_{k}\}\rangle\mid^{2})]^{\frac{1}{n-1}}, \notag
\end{align}
with the minimum is $[\prod\limits_{j>k}(\mid\frac{1}{2}\langle[A_{j}, A_{k}]\rangle\mid^{2}+\mid\frac{1}{2}\langle\{\widehat{A}_{j}, \widehat{A}_{k}\}\rangle\mid^{2})]^{\frac{1}{n-1}}$.
Meanwhile if the MEPS $|\psi^{\perp}_{jk}\rangle$ vary, Eq. (\ref{e:amended3})
provides a family of mutually exclusive uncertainty relations for arbitrary $n$ observables with   (\ref{e:multi4})
as the lower bound.$\blacksquare$

\section*{Acknowledgments}

We thank Jian Wang, Yinshan Chang, Xianqing Li-Jost and Shao-Ming Fei for
fruitful discussions.
The work is supported by
National Natural Science Foundation of China (Grants Nos. 11271138 and 11531004), China Scholarship Council
and Simons Foundation Grant No. 198129.

\section*{Author contributions statement}

Y. X. and N. J. analyzed and wrote the manuscript.

\section*{Additional information}

\textbf{Competing financial interests:} The authors declare no competing financial interests.
\vspace{1ex}

\leftline{\textbf{How to cite this article:} Xiao, Y. and Jing, N. Mutually exclusive uncertainty relations, Sci. Reports, 2016}

\end{document}